\documentclass[12pt]{article}
\usepackage{graphicx}
  \usepackage{color}
\usepackage{epsfig}
\usepackage{amsmath}
\usepackage{hyperref}
\begin{document}

\begin{centering}
\title{  
Should one (be allowed to) \\ replace the Cippolini's?
\vskip0.5cm 
}
\vskip0.5cm
\end{centering}
\maketitle

\author{\;\;\;\;\;\;\;\;\;\;\;\;\;\;\;\;\;\;\;\;\;\;\;\;\;\;\;\;\;\;\;\;\;\;\;\;\;\;\;\;Marcel AUSLOOS$^{a,b,c,d}$  
\\ \\$^a$ School of Business, University of Leicester,\\
Brookfield, 
Leicester, LE2 1RQ, UK\\   e-mail: ma683@leicester.ac.uk
 \\ $^b$ Group of Researchers Applying Physics in Economy and Sociology \\(GRAPES),  Beauvallon, rue de la Belle Jardini\`ere, 483/0021\\ Sart Tilman, B-4031, Li\`ege Angleur, Belgium  \\
  e-mail: marcel.ausloos@uliege.be
  \\$^c$   Universitatea Babeș-Bolyai, \\ Str. Mihail Kogălniceanu nr. 1, 400084, Cluj-Napoca,  Romania
  \\  
  $^d$ Department of Statistics and Econometrics,  \\ Bucharest University of Economic Studies,  15-17 Dorobanti Avenue, \\ District 1, 010552, Bucharest, Romania, \\  e-mail: marcel.ausloos@ase.ro
}

\newpage

\begin{abstract}
One examines and discusses proposals on whether riders could be replaced in a team during multi-stage races,  and how much a team final time at the end of the race would change (be "adjusted")  if only  the riders having completed the race are taken into account for ranking teams.

A few results of the two main multi-stage races, the men Tour de France and the Giro d'Italia, are used as case studies. The impact of disqualification later on,   due to doping,  much after the end of such a race, is also examined in the case of two Tour de France.
 
   The statistical discussion is based on the Kendall-$\tau$ coefficients  for comparing team  ranks at the end of these multi-stages races cases. One observes  {\color{black}that}  there are significant differences in the results of the discussed measures. It is shown that there is much variety in results significance, whence demonstrating many interests of the "adjusted indicators".
Moreover, it is argued that the "adjusted" rank indicator would promote more competitive   {\color{black}and more attractive daily} stages and lead to more valuable race  {\color{black} management}. 
 
\end{abstract}

\bigskip
   {\bf Keywords:   } 
  
    dynamics of social systems;
     hierarchy selection;
   Kendall $\tau$ rank-order correlation coefficient;
   multi-stage  cycling races; 
    teams ranking.
   
\maketitle

\newpage
  
 
 \section{Introduction}\label{Introduction}

Cyclist races are known to be athlete competitions with drastic specificities. A rider wins a race but the success can be  due to his/her teammates help. In this respect, athletes of different teams can collaborate together against other teams,  including   {\color{black}\textit{a priori}}  rival competing riders  (Albert, 1991; Hoenigman et al., 2011; Netland et al., 2012; Mignot, 2015a, 2015b, 2022;  Landkammer et al., 2019).  
 Incidentally, one can consider that cyclist team racing is an experimental process mapping human social class structures (Dos Santos, 1970; Sonubi et al., 2016;  Hadzibeganovic et al., 2018; Nikolova and Vitanov, 2020; Krawczyk and Kułakowski, 2020;  Stefani et al.,  2021; Pham et al., 2022).  

In professional cyclism, in particular in multi-stage races, team managers organize their team according to riders skills: sprinters, mountain climbers, punchers, etc. It is known that  such races  often start with several "flat stages'' leading to a final sprint by bunches. Later on, sprinters might appear to be   "less valuable" in mountain or time trial stages. Yet, several sprinters are helped to be maintained in some "grupetto", during mountain stages, in order to drag these riders as potential winners of the last, very prestigious, stage.

 Suggestions have been recently raised on   whether  one could  replace riders in a team during multi-stage races  (Ausloos, 2024; Unzué, 2024).  
 Ausloos  mocked the idea{\color{black}, after} Unzué had proposed that some restriction might be potentially  introduced:  e.g., the replacement by a substitute should be allowed only during the first week of the race and only in case of a rider sickness or fall.  Notice that there is no discussion {\color{black}yet} on (jury {\color{black}{des commissaires making}}) exceptions to  UCI   {\color{black}(Union Cycliste Internationale; \url{https://fr.uci.org/})} rules, i.e., no replacement of riders arriving (once, or several times!)  out of time\footnote{\color{black}{One might remember a rider, penalized (on stage 14) for doping,  arriving "out-of-delay", on the     19th stage, but at once re-installed by the "jury des commissaires", immediately winning the  20th stage, during Tour de France 1969!}}.
 
   At this level, it seems that an argument for "modernizing rules in cyclist multi-stage races" is that an athlete replacement is allowed during  competitions in many other main sports: teams are permitted to substitute players in basketball, football, soccer, ice or not hockey, rugby, handball, baseball, water polo, tennis, etc.,  during a game
    Such replacements are said to be part of a team coach strategy. Sometimes, once removed, players are not permitted to return to the game in any capacity.
 
 Thus, along such a  {\color{black} concern},  one could also imagine a new rule allowing to replace a player during a double tennis game.  "How often?" might be a subsequent question, since in basketball, football, hockey, handball,  {\color{black} etc.}, it is common to substitute players at will.   {\color{black} In contrast, in the WRC (World Rally Championship; \url{https://www.fia.com/events/world-rally-championship/season-2024/fia-world-rally-champfastestionship-2024}), rules allow  part-time drivers to compete,  - thereby leading to much advantage to these drivers and to their car manufacturer,  due to the subsequently imposed late starting order position for the stages.}
 
Furthermore, a hard to answer question, which does not seem to have been raised by opponents to the Unzué suggestion, concerns the ranking of riders, whence teams, during and at the end of the race. 

{\color{black}{ Indeed,  present UCI rules (according to Chapter VI {\color{black} \footnote{ \url{https://assets.ctfassets.net/761l7gh5x5an/5zPSLYnROebDjMLDWUnLsM/3a2577b206aabcac57ee0471991c887b/2-ROA-20240701-F.pdf}}}), obtain each team final ranking, after summing up the final time of the fastest 3 riders  of the team after each stage. Thus,}} 
 should one sum-up the time (and sprint points) of different riders, -  those of the starting rider and of the substitute(s)? Being more specific for illustrating the argument, recall that it is well  known that M. Cipollini won 
4 consecutive  (“flat'') stages,  in the Tour de France 1999, but abandoned the race on the next rest day, when stages were reaching mountain climbing features\footnote{For completeness, recall that he finished  3rd on the 2nd stage and 10th in the 3rd stage, whence further contributing to the team finishing time for such stages.}. If he had been replaced, say by some better climber rider, should one sum up the times of both riders thereafter?  Should the points obtained by Cipollini (for the green jersey) be transferred to the substitute?
 
 {\color{black}{Within such a line of thought,}} another fundamental question appears: how does one thereafter measure the team performance, - its ranking? That seems very complicated. Indeed, Cipollini's times contributed to the   “team  time" measure on each stage, and at the "final team time"  of the race,  both through his win or when he was in the fastest 3 riders of the team (SAE) for the relevant stage.  {\color{black}{Thus, the}} present UCI rule (according to the above  mentioned Chapter VI), i.e.,  {\color{black}{obtaining the team final time, from the final time of the fastest 3 riders of a given team after each stage}}, irrespective of the fact that a rider might not finish the race, is  much  debatable (Ausloos, 2024).   
  
 In view of these two considerations, the main research question becomes: should the final (team)  ranking  be the sum of optimal stage results, even if a rider does not finish the race, even though he is substituted?
 
 Many past and present riders, team managers, team sponsors, sportscasters, fans, are disagreeing, scorning upon the Unzué suggestion.  a main point concerns the special ingredient characterizing   (long or short) multi-stage races: the rider, to be a 3-week specialist,  knows well that he is going to suffer on various day stage profiles, and be rewarded for his(her) specific effort.  Subsequently, it is known that some riders are aware that they will not compete till the end of the race. To quote A. Valverde "{\it in the last week, when you have to give everything, the people who are really fighting for the GC [general classification] have something extra... }"  (Rendell, 2023).
 
 Therefore, in the following, 
  one examines  how much a team final times at the end of the race would change, with respect to the present UCI rule,  if only  the times of riders having completed each stage of the race are taken into account for ranking teams.

A few results of the two main multi-stage races, the men Tour de France (TdF) and the Giro d'Italia (GdI), (O'Brien, 2017) are used as case studies providing evidences. Of interest, for  the former, one can select the 1999 won by L. Armstrong, - later disqualified (in 2012), when Cipollini withdrew, but also the 1998, won by M. Pantani during which many riders withdrew or were disqualified for doping\footnote{For the list see
 \url{https://fr.wikipedia.org/wiki/Tour_de_France_1998}; see also
  \url{https://www.europe1.fr/sport/Tour-de-France-l-edition-1998-revue-et-corrigee-sans-les-dopes-579604}}. Two more recent cases the 2022 and the 2023 both won  by J. Vingegaard, but in the former M. Quintana (who finished 6th) was disqualified months after the end of the race.

Thus, the subsequent impact of disqualification, later on  much after the end of  a race,  due to doping, is examined in  two cases: TdF 1998 with Armstrong and the TdF 2022 with Quintana. A comparison of "results" taking into account final times as if the disqualification had not occurred is found in an Appendix.

Notice that in  2013, it was announced that tests carried out by the AFLD  {\color{black}(Agence Française de Lutte contre le Dopage; \url{https://www.afld.fr/}}) in 2004 demonstrated the use of erythropoetin (EPO) by Cipollini and Pantini (and several others, like Ullrich and Jullich, thus the top 3 fastest riders) during the 1998 TdF, yet Pantani is still listed as the winner.

Concerning GdI, one examines the  2022 and 2023  years;  
in the latter (2023)  a pertinent example of ranking configurations  due to the no replacement rule occurred: a huge set of riders abandoned the (3 week long) Giro d'Italia, - because of Covid constraints, after about one week.  This   {\color{black}epidemics} decimated the Soudal-Quick Step team which finished the race with only 2 riders. Several  riders, e.g., F. Ganna, R. Evenepoel, S. Gandin, A. Vendrame,  had been  implied in the team first week standings, whence had much implication on the final (time) team ranking. This case confirms the arguments sustaining the  aim and discussion of this study, i.e., the importance of taking the “value" of the only finishing riders in measures, and the quasi non-sense of arguing about replacing riders during a multi-stage race.
 
 The rest of the paper content goes as follows. The methodology used in tackling the research questions is found  in Section \ref{Methodology}. The ranking indicators are defined based on the UCI rules and "adjusted" on the constraint that concerned cyclists must finish the multi-stage race.
  
  The data used in this study, taken from the mentioned races, are briefly described in Section \ref{Data}. The results are presented in Section \ref{Results} and briefly commented upon.

    In Section \ref{Discussion},  pertaining to some   “analysis",    the statistical discussion of results is based on the Kendall-$\tau$ coefficients classically used for comparing ranks in equal size  lists (Kendall, 1938; Abdi, 2007; Puka, 2011){\color{black}; the more so in studies on ranking in competitive sports, e.g., Barrow et al. (2013); Soto-Valero et al. (2019); Ochieng et al. (2022).}  
    
  {\color{black}  No need to say that other techniques are of value in comparing hierarchies; for completeness and comparison, see, e.g., Corvalan, 2018; Csató, 2020, 2023; Ausloos, 2014, 2023, 2024; D'Urso et al., 2023; Ficcadenti et al., 2023; Vernon-Carter et al., 2023), beside other works by the author and co-workers (Ausloos et al., 2014a, 2014b, 2024).} 
    
     {\color{black}In brief, it is  going to be readily } observed that there are significant differences between the results of both discussed indic{\color{black}a}tors. Thus, it is shown that the great variety in results and their high statistical significance demonstrate many possible  interests of the "adjusted indicators", i.e., from sporting or/and economic focuses. 

In Section \ref{Conclusions}, the latter  contains {\it a posteriori} arguments in favour of the new  indicators,  with suggestions for further research and applications in organizations confronted to a selection process based on a hierarchy list. 

Moreover, it is concluded in Section \ref{Conclusions} that the ranking indicators would promote more competitive races, not only till the end of the race, but also until the end of each stage. 
 

 \section{Methodology} \label{Methodology}
   
  There are many papers published on   “ranking teams",   e.g., among pioneers, Sinuany-Stern (1988), Churilov and Flitman (2006), and  Dadelo et al. (2014), -  although many less than on athletes rankings. 
   
   Related to the present considerations, one may point to work on ranking teams globalizing over several races (Rogge et al., 2012). Much complexity is known to exist, sometimes over-complexifying through weighing indicators or ranks; 
 for a recent literature review of contemporary interest on cycling, see Van Bulck et al. (2023). 
  
  Let a few notations and practical steps be so introduced for this paper. 
  
  \subsection{Classical UCI team ranking method}\label{UCIdefs}
  At the end of a stage, the teams are ranked according to the aggregated finishing time of the fastest 3 riders of a team for that stage, -  excluding all so called bonus time. 
   Indeed, according to rule 2.6.014  (and 2.6.021) 
"Bonuses are only taken into consideration for the \underline{individual} general classification."
That sum is cumulated after each stage, such that at the end of the multi-stage race,  each final team time  results from the sum of all stage times of a team, irrespectively of the involved riders (rule 2.6.016).
 
 Practically, call the  team ${(\#)} $ 3 fastest riders, $i=1,2,3$, in their arriving  order  for a stage $s$.
 In mathematical terms, one calculates the   “team (finishing) time  for the  stage $s$" as 
 \begin{equation}\label{tsteameq}
t_s^{(\#)}  = \Sigma_{i=1}^{3} \;\;  t^{(\#)}_{i,s}\;.
\end{equation}
where $ t^{(\#)}_{i,s}$ the finishing time of rider $i$ for that stage.

 At the end of  a  $L$ stages race, a team $  {(\#)} $   “finishing time" is  $T_L^{(\#)}$ resulting from the sum of each stage   “team time":
\begin{equation}\label{Tteameq}
  T_L^{(\#)}  = \Sigma_{s=1}^{L} \;\; t_s^{(\#)}  \;,
\end{equation}
leading to  finally rank the teams at the end of the specific multi-stage race.
 
\subsection{"Adjusted" (or "real") Team  Final  Time}    \label{finalteamtime} 

Since   one can argue that riders should not give up too early in the race,  but should keep up in order to be among their team 3   “best" ( = fastest) riders, whatever their skill,  it seems bizarre that the   “final team ranking" {\color{black} can be partially } based on missing riders at the end of a race, as UCI rules. Thus, one can propose an alternative to $T_L^{(\#)}$.  

 As explained here above, one can  calculate  the  thereafter called   “adjusted team final time", $ A_L^{(\#)} $, defined as follows 
 
 \begin{equation}\label{Ateameq}
  A_L^{(\#)}  = \Sigma_{j=1}^{3} \; \; t_{j,L}^{(\#)}
\end{equation}  
where, in Eq.(\ref{Ateameq}),  $j$ = 1, 2, 3 refers to the   “3 best", whence fastest,  riders of the team ${(\#)}$  {\it having  completed all}  $L$ stages.  

Some possibility exists to   generalize the  $A_L^{(\#)}$   concept and its application; see the conclusion Section on    “further research" suggestions.

 Let it be again emphasized that these 3   “$j$" riders  {\it might} be quite different from the various 3   “$i$" riders having contributed to any  $t_s^{(\#)}$, 
whence to   $T_L^{(\#)}$, - i.e.,  according to UCI way of counting.

 \section{Data} \label{Data}  
 
 \begin{table}\begin{center}  \fontsize{8pt}{8pt}\selectfont
\begin{tabular}{|c||c|c||c|c||c|c|} 
 \hline 
&\multicolumn{4}{|c||}{TdF} &\multicolumn{2}{|c|}{GdI}   \\	\hline
 &	$1998$ &	$1999(^*)$ 	&	$2022(^{**})$	& $2023$ & $2022$ 	&   $2023$    \\ 
  \hline	\hline			
$L$	&	21(+P)	&	20(+P)	&	21	&	21		&	21	& 21\\	
$n$	&	9	&	9	&	8	&	8		&	8	&	8	\\	
$M$	&	21	&	20	&	22	&	22		&	22	&	22	\\	
starters ($N_0$)	&	189	&	180	&	176	&	176		&	176	&	176	\\	
finishers ($N_L$)	&	96	&	141(*)	&	135($^{**}$)	&	150		&	149	&	125	\\	
rider winner	&	Pantani	&	Armstrong(*)	&	Vingegaard	&	Vingegaard		&	Hindley	&	Roglic	\\	
team winner	&	COF	&	BAN	&	IGD	&	TJV	&	TBV	&	TBV\ 
\\ \hline				\end{tabular}
\caption{Characteristics values of the multistage races considered in the main text; $L$:  the number of stages; $n$: the number of  starting riders in a team ($\#$); $M$: the number of teams considered by the organizers; thus there are $nM\equiv N_0$ starting riders; there are $N_L$  riders who finished the race.  The rider winner and the team winner, according to UCI are recalled; (+P) means that a prologue stage took place on the first day;  (*) L. Armstrong was later disqualified, thus only $ N_L=140$ riders are ranked now; ($^{**}$)  N. Quintana was later disqualified, whence only $ N_L=134$ riders are ranked nowadays.} 
\label{TableTdFGdILnM}
 \end{center}
 \end{table}
 
  Both for the Tour de France and the Giro d'Italia,
 the complete list of riders at the end of each stage can be easily downloaded and stored according to their  arrival time and/or place from official or not freely available websites. One should consult, for example the official websites
 
 \url{https://www.letour.fr/en/rankings/stage-21}, official but not  user friendly
 
 \url{https://www.giroditalia.it/fr/classifiche/}
 
 or media websites like
 
  \url{https://www.procyclingstats.com/race/tour-de-france/1998/gc} 
 or
 
  \url{https://www.procyclingstats.com/race/giro-d-italia/2023/gc}, and adapt the year and classification of interest  (riders or teams) accordingly.  Several Wikipedia and Dicodusport (and Dicodutour) are also websites of interest.
  
 Technically speaking, it is sometimes convenient to organize and to store the lists according to the bib   {\color{black}(or plate number attributed at the beginning of the race by organizers)} of each rider. Even though some algorithm can be invented, some summations are more conveniently done manually.
 
 A minor detail about data reliability and exactness may be briefly discussed for completeness: the sum in Eq.(\ref{Ateameq}) for $A_L$ has to be taken over the final time of riders. This rider time on web sites takes into account possible bonuses. However,  UCI rules, whence the time $T_L$, Eq.(\ref{Tteameq}), do not include such time bonus.
 Thus, in order to compare both indicators, one should prefer that $A_L$ does not include such bonus times.  However, the quoted time of riders in the general classification also includes bonus occurring due to intermediary  sprints. This is not usually given in available summary tables. Sometimes, the occurrences are not even mentioned. One might be able to obtain  such time bonus after comparing two successive stages with respect to the published general classification. An extra complication in the "final time correction process" pertains to the fact that such time bonuses do not occur in "mountain stages".  Therefore it is understandable that to sort out the various sources of time bonus becomes a very lengthy, much time consuming process. Moreover recall that the relevant bonuses for the mentioned sums should only concern the 3 fastest  finishing riders of a team. There are very rarely those having gotten a bonus at such intermediary sprints or at final sprints at the end of "flat stages". As an argument for pursuing toward such a fine precision stems from the observation that the difference between the final including or excluding time bonus is usually small, less than 1 minute; I found only two "large cases", 1:01 and 1:28 minutes\footnote {\color{black}  in brief:  on Tour de France 2023, for UAE  51" (Pogačar) + 10" (Yates, A.);  and on Tour de France 2022 for Jumbo-Visma : 32"  (Vingegaard) + 42" (Van Aert) +14" (Laporte), all these riders having brought time contributions  to the team aggregated time on many stages..}
 . This does not lead to any change in the ranking order of teams, - for the cases hereby studied. Therefore, after many attempts at a clearing work, it has been decided to report and to consider $A_L$ values which are strictly the sums of the riders final time quoted in the general classification.

 Table \ref{TableTdFGdILnM}  one reports characteristics of the 6 discussed multi-stage races: the number of stages is $L$, the number of  starting riders in a team ($\#$)  is $n$; $M$ teams are considered by the organizers; thus there are $nM\equiv N_0$ starting riders.  The number of starting ($N_0$) and finishing ($N_L$) riders, the winner, and the team winner, - according to UCI are given.
  For each team, the official UCI code is hereby used, - for shortening the writing and avoiding undue publicity claims.

              \begin{table}\begin{center} 
\begin{tabular}{|c||c|c||c||c|c|c|c|c|c|c|} \hline 
 rank	&$T_L^{(\#)}$	&team      & &$A_L^{(\#)}$ &     team  
  \\ \hline   \hline
1	&	278:29:58	&	COF&=&	278:57:55	&	COF\\	
2	&	278:59:07	&	CSO&x&	279:29:14	&	TEL\\	
3	&	279:11:38	&	USP&x&	280:09:51	&	LOT\\	
4	&	279:15:59	&	TEL&x&	280:26:53	&	USP\\	
5	&	279:34:12	&	LOT&x&	280:32:22	&	MAP \\	
6	&	279:36:30	&	PTK&x&	280:47:28	&	CSO\\	
7	&	280:16:18	&	RAB&x&	280:54:31	&	PTK\\	
8	&	280:29:51	&	MAP &x&	281:13:54	&	RAB\\	
9	&	280:33:30	&	BIG&x&	281:20:38	&	MER\\	
10	&	280:53:02	&	MER&x&	281:33:41	&	FDJ\\	
11	&	280:53:33	&	FDJ&x&	281:44:11	&	BIG\\	
12	&	281:55:21	&	SAE&=&	282:41:21	&	SAE\\	
13	&	282:03:09	&	GAN&=&	282:50:17	&	GAN\\	
14	&	282:35:09	&	ASI&=&	283:45:43	&	ASI\\	
\hline
15	&	-	&	FES	&&	-	&	FES	\\	
16	&	-	&	ONC	&		&-&	ONC	\\	
17	&	-	&	BAN	&		&-&	BAN	\\	
18	&	-	&	TVM	& &-		&	TVM	\\	
19	&	-	&	KEL	& &-		&	KEL	\\	
20	&	-	&	RIS	& &-		&	RIS	\\	
21	&	-	&	VIT	&		&-&	VIT	\\	
\hline	\end{tabular}
\caption{Resulting time ranking of teams  {\it at the end of the 1998  TdF}, according to the $T_L^{(\#)}$ and  $A_L^{(\#)}$   indicators,  defined in Eq.(\ref{Tteameq}) and  Eq.(\ref{Ateameq}), respectively; thus, on one hand, from the (usual, UCI) sum of the finishing time of the   “fastest" 3 riders of the team  {\it after each stage}, and, on the other hand,  hereby  “adjusted" in order to be only taking into account those riders who did finish the whole race,  respectively.  The central column emphasizes whether the ranking is the same (=) or not (x) in both indicators. Recall that 7 teams were excluded for doping.}\label{Table1998TdF} 
 \end{center}
 \end{table}  
 
         \begin{table}\begin{center} 
\begin{tabular}{|c||c|c||c||c|c|c|c|c|c|c|} \hline 
 rank	&$T_L^{(\#)}$	&team      & &$A_L^{(\#)}$ &     team  
  \\ \hline   \hline
1	&	275:05:21	&	BAN&	x	&	275:43:26	&	ONC\\	
2	&	275:13:37	&	ONC&	x	&	276:08:32	&	MQS\\	
3	&	275:21:34	&	FES&	x	&	276:21:09	&	VIT\\		
4	&	275:29:09	&	KEL	&	=	&	276:23:27	&	KEL	\\	
5	&	275:29:34	&	MQS&	x	&	276:27:06	&	TEL\\	
6	&	275:46:21	&	TEL&	x	&	276:44:57	&	BAN\\	
7	&	275:48:05	&	VIT&		x	&	277:00:41	&	FES\\	
8	&	276:02:34	&	USP&	x	&	277:00:56	&	SAE\\	
9	&	276:03:23	&	COF&	x	&	277:03:05	&	LOT\\	
10	&	276:14:23	&	LOT&	x	&	277:18:53	&	COF\\	
11	&	276:31:50	&	CSO&	=	&	277:23:36	&	CSO\\	
12	&	276:51:58	&	SAE&	x	&	278:08:52	&	USP\\	
13	&	278:00:27	&	PTK&	x	&	278:42:03	&	FDJ\\	
14	&	278:06:32	&	FDJ&	x	&	279:03:40	&	MER\\	
15	&	278:28:34	&	MER&	x	&	279:05:03	&	PTK\\	
16	&	279:05:39	&	BIG&	x	&	279:36:19	&	C.A\\		
17	&	279:12:45	&	C.A&		x	&	280:04:39	&	BIG\\		
18	&	279:19:34	&	LAM&	=	&	280:30:13	&	LAM\\	
19	&	279:56:14	&	RAB&	=	&	280:46:21	&	RAB\\	
20	&	280:16:23	&	CTA&	=	&	282:16:30	&	CTA\\	
  \hline	\end{tabular}
\caption{Resulting time ranking  of teams  {\it at the end of the 1999  TdF},  ($N_L =140$) 
according to the $T_L^{(\#)}$ and  $A_L^{(\#)}$   indicators,  defined in Eq.(\ref{Tteameq}) and  Eq.(\ref{Ateameq}), respectively; thus, on one hand, from the (usual, UCI) sum of the finishing time of the   “fastest" 3 riders of the team  {\it after each stage}, and, on the other hand,  hereby  “adjusted" in order to be only taking into account those riders who did finish the whole race,  respectively.  The central column emphasizes whether the ranking is the same (=) or not (x) in both indicators.}\label{Table1999TdFnoArmstrong} 
 \end{center}
 \end{table}  
 
      \begin{table}\begin{center} 
\begin{tabular}{|c||c|c||c||c|c|c|c|c|c|c|} \hline 
 rank	&$T_L^{(\#)}$	&team      & &$A_L^{(\#)}$ &     team  
    	\\\hline \hline
1	&	239:03:03	&	IGD&	=	&	240:13:45	&	IGD\\		
2	&	239:40:36	&	GFC&	=	&	240:20:17	&	GFC\\	
3	&	239:47:57	&	TJV&	=	&	241:19:52	&	TJV\\		
4	&	240:51:48	&	BOH&	=	&	242:36:44	&	BOH\\	
5	&	241:14:25	&	MOV&	x	&	243:21:28	&	DSM\\	
6	&	241:22:57	&	UAD&	x	&	243:33:42	&	AST\\	
7	&	242:01:35	&	TBV&	=	&	243:45:06	&	TBV\\	
8	&	242:29:11	&	DSM&	x	&	243:50:09	&	EFE\\	
9	&	242:59:54	&	ARK&	x	&	244:06:03	&	MOV\\	
10	&	243:02:03	&	AST&	x	&	244:24:47	&	IWG\\	
11	&	243:05:48	&	EFE&	x	&	245:14:59	&	UAD\\	
12	&	243:09:27	&	IPT&		x	&	246:08:10	&	TFS\\	
13	&	243:20:42	&	TFS&	x	&	246:53:22	&	ACT\\	
14	&	243:24:40	&	IWG&	x	&	246:55:09	&	COF\\	
15	&	244:31:23	&	ACT&	x	&	247:44:49	&	ARK\\	
16	&	244:54:33	&	COF&	x	&	247:53:06	&	IPT\\		
17	&	245:40:01	&	BBK&	=	&	248:02:04	&	BBK\\	
18	&	247:03:40	&	BEX&	=	&	248:21:53	&	BEX\\	
19	&	247:14:16	&	TEN&	=	&	248:37:14	&	TEN\\	
20	&	248:22:12	&	ADC&	=	&	249:33:21&	ADC\\	
21	&	248:38:47	&	LTS&	=	&	250:28:24	&	LTS\\		
22	&	249:46:16	&	QST&	=	&	251:22:16	&	QST\\	
  \hline	\end{tabular}
\caption{Resulting time ranking of teams  {\it at the end of the 2022  TdF} ($N_L =134$) 
according to the $T_L^{(\#)}$ and  $A_L^{(\#)}$   indicators,  defined in Eq.(\ref{Tteameq}) and  Eq.(\ref{Ateameq}), respectively; thus, on one hand, from the (usual, UCI) sum of the finishing time of the   “fastest" 3 riders of the team  {\it after each stage}, and, on the other hand,  hereby  “adjusted" in order to be only taking into account those riders who did finish the whole race,  respectively.  The central column emphasizes whether the ranking is the same (=) or not (x) in both indicators.}\label{Table2022TdFnoQuintana} 
 \end{center}
 \end{table}  
 
         \begin{table}\begin{center} 
\begin{tabular}{|c||c|c||c||c|c|c|c|c|c|c|} \hline 
 rank	&$T_L^{(\#)}$	&team      & &$A_L^{(\#)}$ &     team  
  \\ \hline   \hline			
1	&	247:19:41	&	TJV&	x	&	247:31:40	&	UAD\\	
2	&	247:33:30	&	UAD&	x	&	248:01:24	&	TJV\\		
3	&	247:47:19	&	IGD&	=	&	248:14:52	&	IGD\\		
4	&	247:48:18	&	TBV&	x	&	248:22:27	&	GFC\\	
5	&	248:17:01	&	GFC&	x	&	249:28:33	&	ACT\\	
6	&	249:10:41	&	ACT&	x	&	249:46:20	&	BOH\\	
7	&	249:24:49	&	BOH&	x	&	249:48:58	&	TBV\\	
8	&	250:41:14	&	JAY&	=	&	251:39:29	&	JAY\\		
9	&	251:53:30	&	IPT&		x	&	254:07:00	&	COF\\	
10	&	251:58:07	&	MOV&	x	&	254:08:28	&	ARK\\	
11	&	252:27:17	&	COF&	x	&	254:17:11	&	IPT\\		
12	&	252:32:26	&	ARK&	x	&	254:48:20	&	MOV\\	
13	&	253:16:17	&	DFP&	x	&	254:51:41	&	LTK\\		
14	&	253:25:38	&	LTK&	x	&	255:33:18	&	UXT\\	
15	&	253:41:49	&	UXT&	x	&	255:45:58	&	DFP\\	
16	&	254:16:42	&	ICW&	x	&	256:50:03	&	TEN\\	
17	&	254:33:10	&	TEN&	x	&	256:50:51	&	ICW\\	
18	&	254:35:15	&	AST&	=	&	256:52:24	&	AST\\	
19	&	255:29:23	&	EFE&	x	&	257:31:48	&	LTD\\	
20	&	256:36:01	&	SOQ&	=	&	257:40:34	&	SOQ\\	
21	&	257:03:28	&	LTD&	x	&	257:53:49	&	EFE\\	
22	&	258:11:39	&	ADC&	=	&	258:59:58	&	ADC\\	
\hline	\end{tabular}
\caption{Resulting time ranking of teams  {\it at the end of the 2023  TdF}, according to the $T_L^{(\#)}$ and  $A_L^{(\#)}$   indicators,  defined in Eq.(\ref{Tteameq}) and  Eq.(\ref{Ateameq}), respectively; thus, on one hand, from the (usual, UCI) sum of the finishing time of the   “fastest" 3 riders of the team  {\it after each stage}, and, on the other hand,  hereby  “adjusted" in order to be only taking into account those riders who did finish the whole race,  respectively.  The central column emphasizes whether the ranking is the same (=) or not (x) in both indicators.}
\label{Table2023TdF} 
 \end{center}
 \end{table}

 
        \begin{table}\begin{center} 
\begin{tabular}{|c||c|c||c||c|c|c|c|c|c|c|} \hline 
 rank	&$T_L^{(\#)}$	&team      & &$A_L^{(\#)}$ &     team  
  \\ \hline   \hline
1	&	259:48:12	&	TBV	&=&	260:10:43	&	TBV	\\	
2	&	259:52:19	&	BOH	&=&	260:28:46	&	BOH	\\	
3	&	261:10:41	&	IGD	&x&	262:10:32	&	IWG	\\	
4	&	261:12:09	&	IWG	&x&	262:44:09	&	IGD	\\	
5	&	262:06:58	&	AST	&=&	262:58:15	&	AST	\\	
6	&	262:09:22	&	TFS	&=&	263:04:34	&	TFS	\\	
7	&	262:28:28	&	TJV	&=&	263:43:47	&	TJV	\\	
8	&	263:09:14	&	UAD	&x&	264:42:54	&	MOV	\\	
9	&	263:18:10	&	BEX	&x&	265:23:58	&	DSM	\\	
10	&	263:27:57	&	MOV	&x&	266:00:59	&	BEX	\\	
11	&	264:18:04	&	DSM	&x&	266:59:21	&	UAD	\\	
12	&	265:05:45	&	ACT	&x&	267:02:47	&	BCF	\\	
13	&	265:29:00	&	EFE	&x&	267:04:33	&	ACT	\\	
14	&	265:37:26	&	BCF	&x&	267:12:41	&	COF	\\	
15	&	265:39:50	&	COF	&x&	267:16:38	&	EFE	\\	
16	&	266:01:30	&	EOK	&=&	267:18:03	&	EOK	\\	
17	&	268:03:33	&	QST	&=&	269:39:06	&	QST	\\	
18	&	268:55:39	&	DRA	&=&	271:44:06	&	DRA	\\	
19	&	270:25:28	&	AFC	&x&	271:50:54	&	LTS	\\	
20	&	271:05:15	&	GFC	&x&	271:59:27	&	AFC	\\	
21	&	271:10:05	&	LTS	&x&	272:22:06	&	GFC	\\	
22	&	273:44:19	&	IPT	&=&	274:52:40	&	IPT	\\	
\hline	\end{tabular}
\caption{Resulting time ranking of teams  {\it at the end of the 2022  Giro}, according to the $T_L^{(\#)}$ and  $A_L^{(\#)}$   indicators,  defined in Eq.(\ref{Tteameq}) and  Eq.(\ref{Ateameq}), respectively; thus, on one hand, from the (usual, UCI) sum of the finishing time of the   “fastest" 3 riders of the team  {\it after each stage}, and, on the other hand,  hereby  “adjusted" in order to be only taking into account those riders who did finish the whole race,  respectively.  The central column emphasizes whether the ranking is the same (=) or not (x) in both indicators.}
\label{Table2022Giro} 
 \end{center}
 \end{table} 

 	   \begin{table}\begin{center} 
\begin{tabular}{|c||c|c||c||c|c||c|c|c|c|c|} \hline 
 rank	&$T_L^{(\#)}$	&team     & &$A_L^{(\#)}$ &     team  
  \\ \hline   \hline	
1	&	256:21:18	&	TBV	&x&	256:42:33	&	IGD	\\	
2	&	256:37:40	&	IGD	&x&	257:18:53	&	TBV	\\	
3	&	256:51:58	&	TJV	&=&	257:28:30	&	TJV	\\	
4	&	257:13:11	&	UAD	&=&	258:38:54	&	UAD	\\	
5	&	257:42:48	&	ACT	&x&	258:45:48	&	JAY	\\	
6	&	257:46:49	&	BOH	&=&	258:50:46	&	BOH	\\	
7	&	257:53:02	&	AST	&x&	258:59:37	&	ACT	\\	
8	&	257:54:09	&	JAY	&x&	259:54:55	&	AST	\\	
9	&	258:13:12	&	IPT	&x&	259:57:49	&	GFC	\\	
10	&	258:40:10	&	EFE	&x&	260:54:25	&	MOV	\\	
11	&	258:51:33	&	GFC	&x&	260:55:33	&	IPT	\\	
12	&	258:58:40	&	MOV	&x&	262:09:55	&	EOK	\\	
13	&	259:42:20	&	EOK	&x&	263:02:32	&	EFE	\\	
14	&	261:34:17	&	ICW	&x&	263:09:22	&	COF	\\	
15	&	261:37:47	&	TSF	&x&	263:11:42	&	ICW	\\	
16	&	262:07:29	&	BCF	&x&	263:57:12	&	TSF	\\	
17	&	262:14:01	&	COF	&x&	264:10:18	&	ARK	\\	
18	&	262:44:35	&	ARK	&x&	264:22:53	&	DSM	\\	
19	&	262:50:33	&	DSM	&x&	264:57:43	&	BCF	\\	
20	&	264:45:58	&	ADC	&=&	266:08:21	&	ADC	\\	
21	&	267:59:00	&	COR	&=&	269:04:23	&	COR	\\	
\hline
22	&		- 	&	SOQ	&=&	-		&	SOQ\\	
\hline	\end{tabular}
\caption{Resulting time ranking of teams  {\it at the end of the 2023  Giro}, according to the $T_L^{(\#)}$ and  $A_L^{(\#)}$   indicators,  defined in Eq.(\ref{Tteameq}) and  Eq.(\ref{Ateameq}), respectively; thus, on one hand, from the (usual, UCI) sum of the finishing time of the   “fastest" 3 riders of the team  {\it after each stage}, and, on the other hand,  hereby  “adjusted" in order to be only taking into account those riders who did finish the whole race,  respectively.  The central column emphasizes whether the ranking is the same (=) or not (x) in both indicators. Recall that SOQ had only 2 riders finishing the race.}
\label{Table2023Giro} 
 \end{center}
 \end{table} 
 
  \section{Results}\label{Results}
  
  The pertinent results are given in Tables \ref{Table1998TdF}-\ref{Table2023TdF}, for the TdF's and Tables \ref{Table2022Giro}-\ref{Table2023Giro} for the GdI's. Each table contains the UCI team ranking and the "adjusted" ranking with either official or calculated times as explained here above. The central columns emphasizes when the team ranks are the same in each indicator.  The = sign points to "concordant pairs" in a Kendall-$\tau$  coefficient analysis; see below. 
  
  It can be at once noticed that  $A_L^{(\#)}$ is always greater than $T_L^{(\#)}$. This can be easily understood : the 3 riders best final times are usually not those having contributed to $T_L^{(\#)}$,  in particular at the beginning of the race.
  
  Moreover,  it can be observed that the ranks are much scrambled in all cases.

  \subsection{Tour de France}
  
\begin{figure}[ht] \centering
\includegraphics[height=11cm,width=12cm]{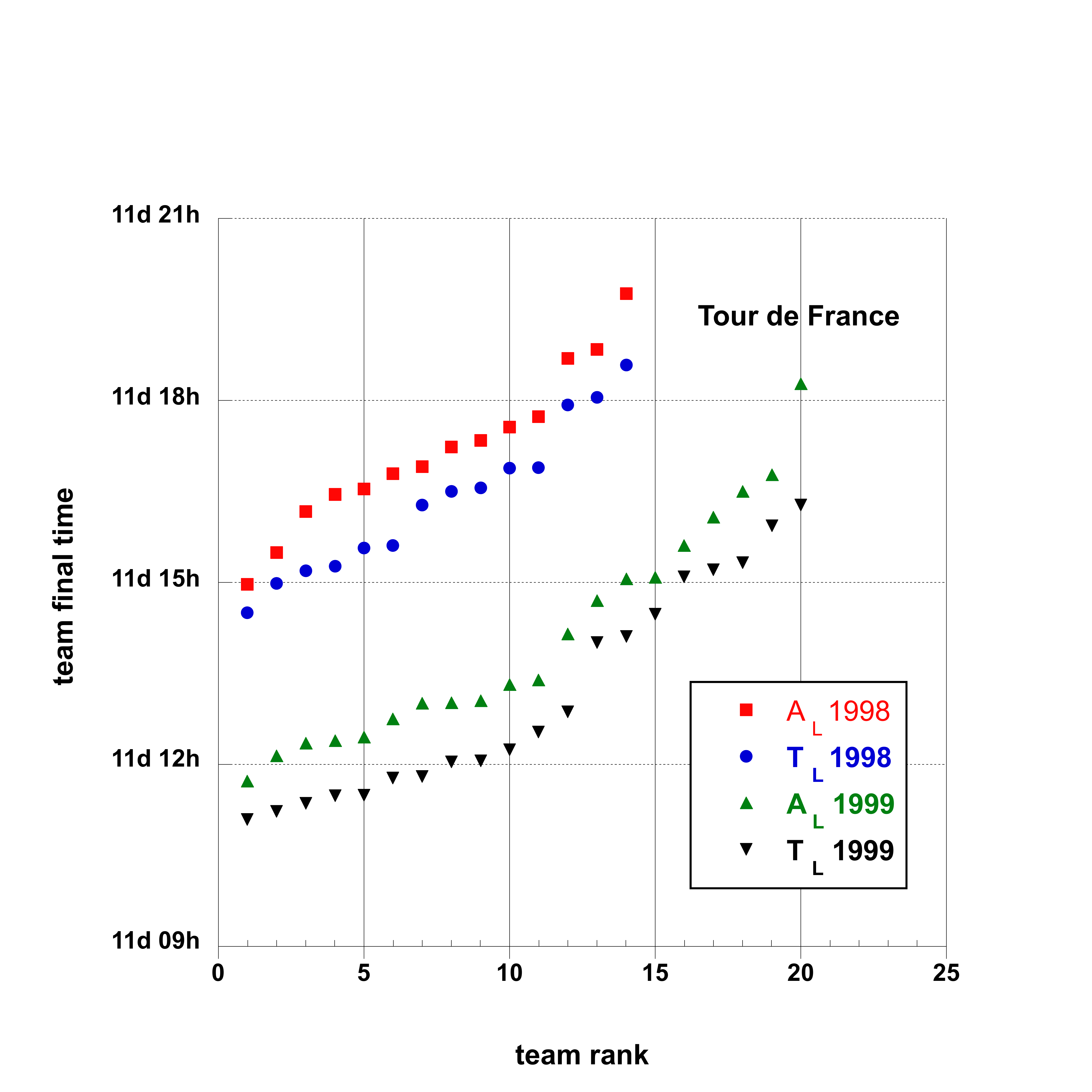}
\caption{Tour de France  1998 and 1999  teams final time  ranked in time increasing order, according  either to UCI ($T_L$)  or to the present study ($A_L$).}
\label{TdF9899}
\end{figure}

\begin{figure}[ht] \centering
\includegraphics[height=11cm,width=12cm]{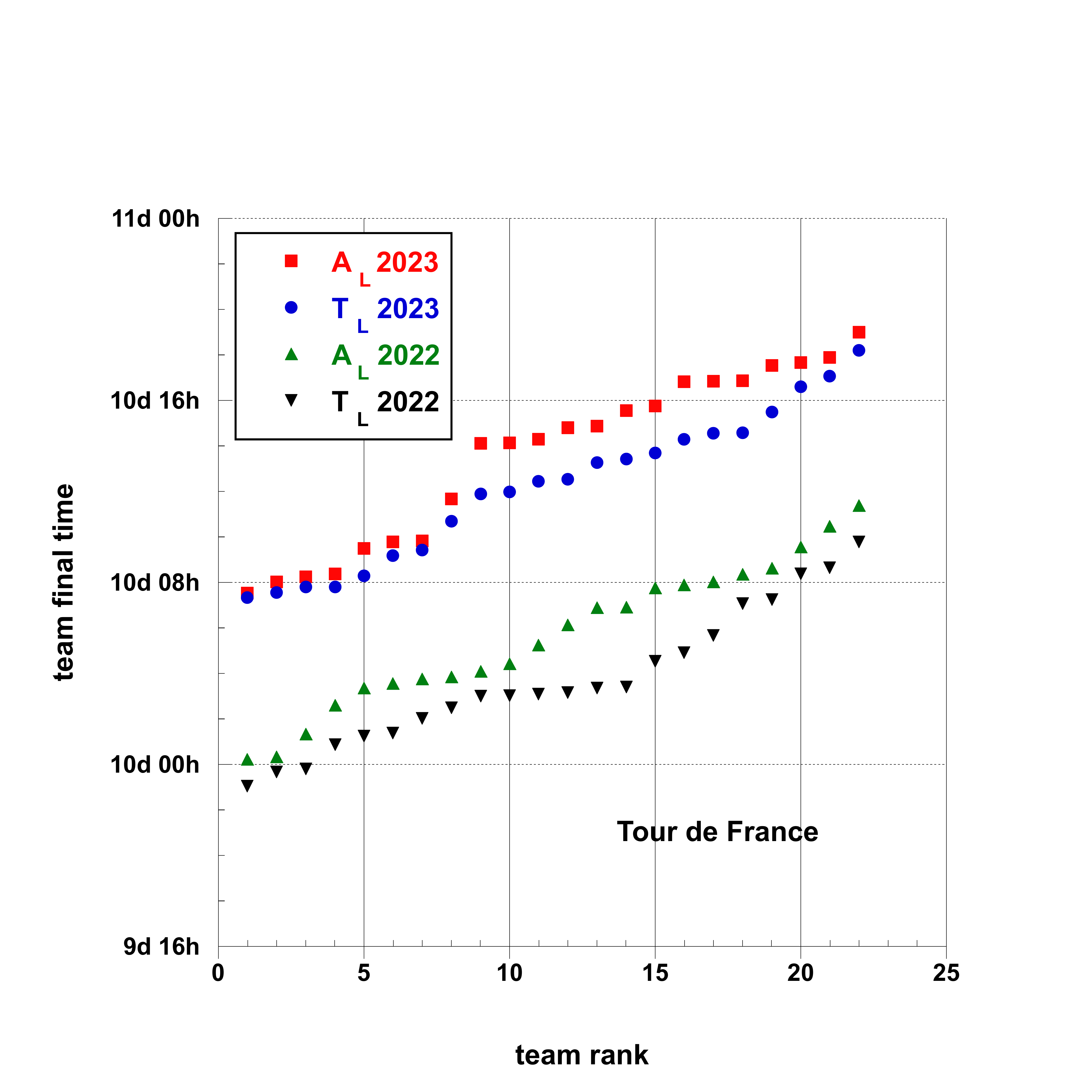}
\caption{Tour de France  2022 and 2023  teams  final time ranked in time increasing order, according either  to UCI ($T_L$)  or  to the present study ($A_L$).}
\label{TdF2223}
\end{figure}

   A display of Tour de France  teams final time,  ranked in time increasing order, according  either to UCI ($T_L$)  or to present study ($A_L$) is presented on Fig. \ref{TdF9899} and Fig. \ref{TdF2223}, respectively for  1998 and 1999  and for 2022 and 2023. Some team clustering is seen; see Section \ref{Discussion} for further discussion.  

    \subsection{Giro d'Italia}
    
\begin{figure}[ht] \centering
\includegraphics[height=11cm,width=12cm]{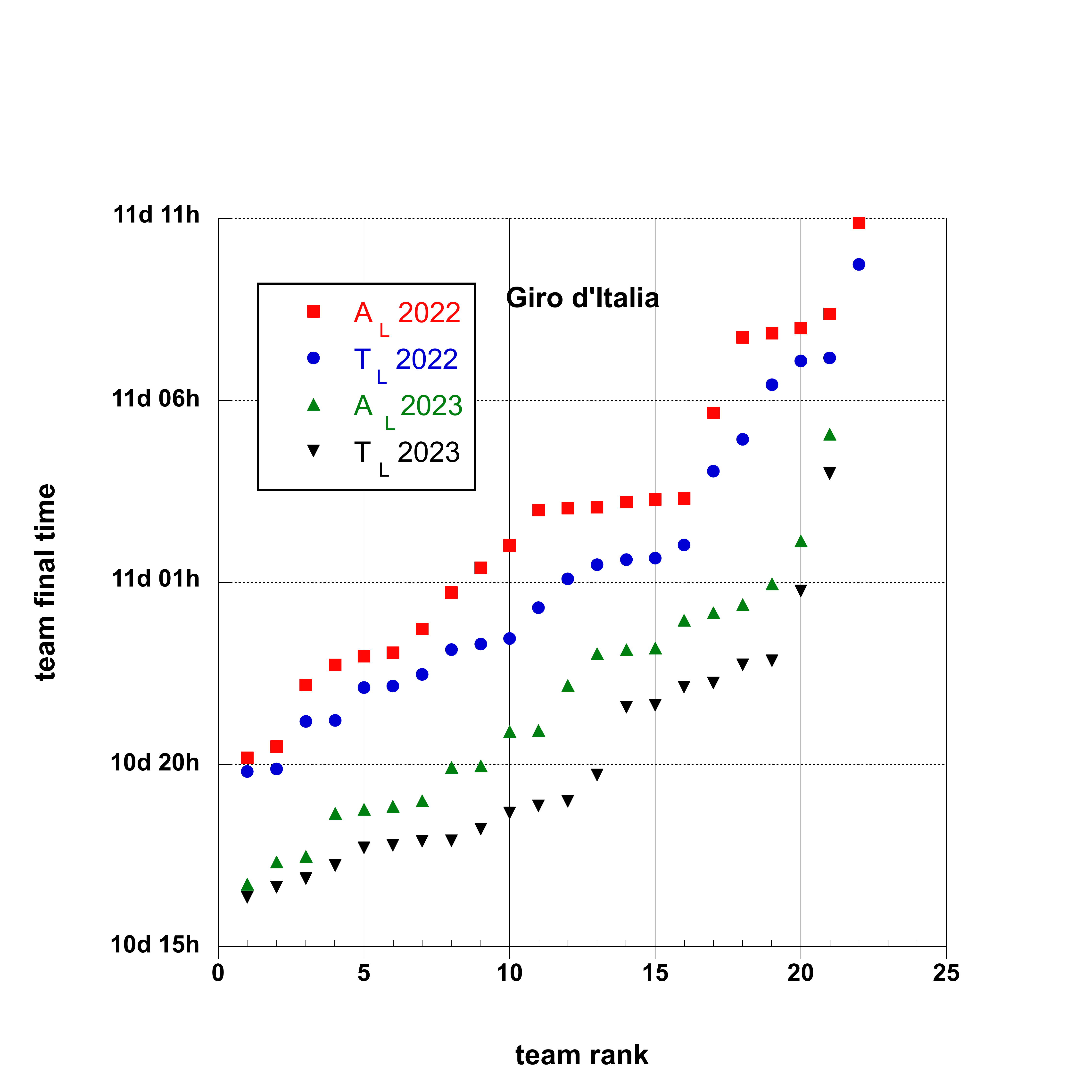}
\caption{Display of Giro d'Italia 2022 and 2023 teams  final time ranked in time increasing order, according either to UCI ($T_L$)  or to  the present study ($A_L$).}
\label{GdI2223}
\end{figure}

    A display of the Giro d'Italia 2022 and 2023 teams  final time, ranked in time increasing order, according either to UCI ($T_L$)  or to the present study ($A_L$) is provided on Fig. \ref{GdI2223}. Again,  some team clustering can be observed; see Section \ref{Discussion} also. 
 
It is vizually remarkable from Tables \ref{Table2022Giro}-\ref{Table2023Giro}  that a 2022 and 2023 ranking equivalence does not exist. The winning team and the second are equivalent; there are only 9 similar ranking; thus 13 differences.
    In 2023, even the winning teams differ; there are only 6 equivalences, when including SOQ.

    \clearpage

 \section{Discussion} \label{Discussion}.

  In order to observe whether the  indicator $A_L$  brings some 
  interests beyond those of $T_L$, i.e., somewhat their framework validity, one may test them as follows. 
 The present statistical discussion is based on the Kendall-$\tau$ coefficients   for comparing team  ranks at the end of these multi-stages races cases; see {\color{black}  a few justifying references in the Introduction section.}
 
 For a pertinent comparison of   indicators, one has  performed a  Kendall-$\tau$ rank-rank correlation test, along 
 \url{https://www.wessa.net/rwasp_kendall.wasp\#output}. The   (free on line) latter website   provides the two-sided $p$-value. Moreover, the    website     provides scatter plots of the $X$ and $Y$ variables and alternatively of their respective ranks. For space saving, these plots are not shown, since they are not carrying any peculiar information of present interest beside that found in examining the Tables or in the displayed graphs.
 
 Conventionally, in such an analysis, the number of concordant pairs is called $C$; that of discordant pairs is $D$. The    “score" $S$  is  equal to  $C-D$.  By definition, the Kendall $\tau = \frac{C-D}{m}$, where the   “denominator"  ($m$)  is the total number of all possible pairs combinations, i.e., $M(M-1)/2$, in the present study. 

{\color{black}{  Notice that there is no {\it ex aequos} in the $T_L$ or $A_L$ list values; nevertheless ties in ranking  are found in other cases, when daily $T_s$ and $A_s$ data are considered. To adapt the Kendall $\tau$ should then be meaningful, including extensions toward the  weighted Kemeny distance (Can, 2014; Csató, 2017; Ausloos, 2024; Cerqueti et al., 2024).} }
 Recall that a positive (negative) $\tau$ indicates a so called high (low) rank correlation. 
  If $\tau \sim 0$, one claims that there is no correlation. 
  
  The results are given in Table \ref{Table5KtauTdFGdI}. In all cases, the results are found to be  correlated, - as somewhat expected, and are statistically significant.
  
        \begin{table}\begin{center}  
\begin{tabular}{|c||c|c||c|c||c|c|} 
 \hline 
&\multicolumn{4}{|c||}{TdF} &\multicolumn{2}{|c|}{GdI}   \\	\hline
 &	$1998$ &	$1999$ 	&	$2022$	& $2023$ & $2022$ 	&   $2023$    \\ 
  \hline	\hline	 
Kendall-$\tau$	&	0.78022	&0.75789 &	0.80952	&	0.87879 &0.90476 & 0.85714	\\	
2-sided p-value	&	0.00013	&	0.00	&	0.00	&	 0.00	& 0.00 & 0.00 \\	
Score	&	71 	&	144 	&	187 	&	203 & 209 & 180 \\	
Denominator	&	91	&	190	&	231	&	231	&231 &  210\\ \hline	 	
\end{tabular}
   \caption{ Illustration of Kendall $\tau$ rank-rank correlation  results between  $T_L^{(\#)}$ and $A_L^{(\#)}$ corresponding to the multi-stage races hereby examined, - taking into account disqualified riders in 1999 and 2022 TdF.
}\label{Table5KtauTdFGdI}
 \end{center}
 \end{table}

Observing the Tables and the Figures, one is rightly tempted to look for team clustering, as observed in other sport or more generally ranking.

One can observe clusters of teams
 \begin{itemize}
\item in TdF 1998: below and above $r_b =12$
\item in TdF 1999: below and above $r_b =12$
\item in TdF 2022: below $r_b =9$ and others below and above $r_c =20$
\item in TdF 2023: below  $r_b =8$ and  others below and above $r_c =19$
\item in GdI 2022: below  $r_b =10$ and others  below and above $r_c =20$
\item in GdI 2023: below  $r_b =10$ and others  below and above $r_c =17$
\end{itemize}
The clusters can be  also guessed from the corresponding  Tables. Recall that the teams respective orders are quite scrambled, but  remaining in the same cluster. 

 \begin{itemize}
\item TdF 1998 :  the best team is the same for $T_L$ and $A_L$; there is much scrambling in the center of ranks; no scrambling for the worst (3) teams;

\item Tdf 1999 : scrambling of the best 3 teams and  much  scrambling in the center of ranks; no scrambling for the worst (3) teams; this leads to the lowest Kendall $\tau$ of this study; the largest difference (5 ranks) in ranking occurs for BAN;

 \item TdF 2022 :  the best (4) teams are the same ones either for $T_L$ or $A_L$; there is much scrambling in the center of ranks; no scrambling for the worst (6) teams; the largest difference (6 ranks) in ranking occurs for ARK;

\item TdF 2023 ; scrambling of the best 2 teams; much scrambling, but mere small swapping, in the center of the ranks,  up to the last ranks;

 \item GdI 2022 :  the best (2) teams are the same ones either for $T_L$ or $A_L$; the 5-th tom 8-th teams are identically ranked in $T_L$ and $A_L$; a same "rank conformity" occurs  for the  16-th to 18-th and the worst team; otherwise   there is much scrambling in the center of ranks; such a  "best conformity" leads to the hugest Kendall $\tau$ value of this study;

 \item GdI 2023 :  the best (2) teams are  swapped in $T_L$ and $A_L$;  but not the 3-rd and 4-th; the internal ranking is much scrambled; the last 2 teams are equally ranked.
 
 \end{itemize}
 
 It is obvious that there is much variety in results significance, whence demonstrating many interests of the "adjusted indicators".

 \section{Conclusions}\label{Conclusions}
 
 It is commonly admitted that cyclist races are won by one rider, but the role of the team is of crucial importance  (Albert, 1991;  Mignot, 2015a, 2015b, 2022).  
  In fact, cyclist races are quite different from other sport competitions, emphasising individual athletes. Even team competitions, like football (soccer), basketball,  hockey, rowing, etc.,  which sometimes have some focus on specific athletes, or even animals (Lessman et al., 2009) have team quality usually derived from (integer) numbers, corresponding to some rank and statistics (Anderson, 2015). 
  {\color{black}  Cyclism,  in particular multi-stage competitions, in contrast rely on time measures.}
 
  Notice that the study  emphasizes an objective
  development and further applications of practical and realistic measures to improve "decision-making" on team values leading to subsequent ordering, whence hierarchy.
  
  The numerical analysis is centered on the final time riders hierarchy.  This can serve as a counter-argument about the recent views about replacing riders during such a type of race.  Observe that some daily adjusted $A_s$ might  be  considered, but this "generalization"  falls outside the present study.
  
  The exploration into the "Adjusted Rankings", facilitated by the examination of  the Kendall’s tau correlations, highlights the potential for alternative standings that deviate significantly from the official (UCI) rankings.   The $A_L$  rankings  demonstrate  substantial shifts in team positions. Whether there is some competition strategy  origin in such a finding can be debated but this is left for fan blogs.
 
  In conclusion, from the present research on, and proposal of, a more real team ranking  indicator, one can observe that  the $A_L$  indicator brings some new quantitative information on the various team's   “values"  at the end of a multi-stage cyclist race. Indeed, it can  be   observed that the  indicators proposes a different team hierarchy, - because only the finishing riders are considered.  
  
  Further studies on past and future strategies are in order. One should remark that the $A_L$ analysis is based on data for which no strategy was {\it a priori} developed; the role of a coach is not introduced. It should be relevant, and even exciting to see how different strategies, beside previously  aimed at $T_L$, will be developed in order to be the most valuable team(s) along the presented operational lines.

Moreover, this points out that  a team excellence overall ranking measure demands a consistent competition by all the riders of a given team.
{\color{black} {No need to emphasize that the  number of riders of a given team finishing a race  ($N_L^{(\#)}$}) is a good measure of a team strength. It could be also used as a weight in defining another indicator.   Notice also  the most recent, whence 2024, Tour de  France which  had very specific features: e.g., ending in Nice, instead of Paris, after again crossing the Alps, during  the 4 final stages, and finishing with a time trial stage, instead of a flat stage usually ending in a glorious sprint of the bunch on Champ Elysées. This led to a remarkable set of riders withdrawals before the  17th stage (out of 21), immediately after the rest day. This  remark (added  when writing the revised version of the paper) usefully serves as a strong argument carrying the present work and a reviewer thoughtful suggestion toward further consideration and studies.  }
  
 It seems that one unduly expects that the  "to be a finishing  rider" accounting  condition should demand more riding action, every day  of the race, till the end, thus fuller and more attractive competition.  That does not prove that rider substitution will lead to a more fair race. {\it A contrario}, through the $A_L$ indicator measure, it seems proved that a more competitive race occurs if riders do not abandon at some stage (and are not replaced).   {\color{black}  This is further elaborated upon as follows below.}

\subsection{Strategic implications}
 {\color{black}  
 One may notice that the requirement about "necessarily finishing" riders might imply team strategies different from the present ones, based on UCI rules. Recall that in the present study the goal is to measure some team "value" through the 3 fastest  riders at the end of the race;   the value is obtained through the general classification rather than the various stages rankings. In some sense, the team must have at least 3 (more, in case of potential "problems")
  predefined  leaders, in contrast to present strategies mainly relying on one leader. However, there might be  some flexibility in the {\it a priori} leadership definition. Indeed, the selected riders for the $A_L$ classification may change during the race. Not all 3  fastest riders at the end of the race need to be the fastest 3 in each stage. Thus, a requirement on a team seems to be a "greater homogeneity" or rider versatility. One may imagine that the richest teams would be thereby favorized. But the team sponsors are already aware of the money needed for building a strongly competitive team. Nevertheless, the team starting composition might be more  differently imagined than as at present, as a function of the type and order of stages in the race competition. Such a point, briefly made in a preceding paragraph about the Tour de France 2024, carries some weight here.
  
  In so doing, one may imagine that the "grupetto scheme" might be  adjusted. The latter usually concerns sprinters grouping each other for tech ical reasons at the end of a stage, in order not to reach the finishing line, "out-of-delay". 
  Along the present study, somewhat  $3M$ riders should compose the leading bunch; these $3M$ riders should frequently be the "leaders". Therefore the present scheme demands much activity from "team leaders" also in "flat stages". But in order to be well served in the leading group, the leaders must be accompanied by other servant riders. Therefore, the scheme seems to be leading to more complex strategies and competitive stages. One might {\it a contrario} argue that each  stage will be dull, since all teams might be willing not to compete at all. This is hardly realistic, knowing that teams and riders have varied aims. Thus, one might forecast clusters of riders through "grupetti", along simple Lotka-Volterra collaboration-competition models (Vitanov et al., 2010; Sonubi  et al., 2016; Stefani et al,. 2021). A lengthy discussion of the variety of strategies for flat, hilly, mountain stages is in order but has better postponed for further studies.
  
Nevertheless, it may be concluded that each daily race would demand a complex implementation  of coevolving interacting strategies, prone to operational research considerations.
  
 }

  \subsection{Economic point of view}
  As other arguments in favour of the  introduction of the $A_L$ ranking measure, one can imagine some interest by sponsors, since the presentation of teams on a podium at the protocol time is the source of a non negligible publicity. 
  
  Notice that  UCI rules only permit 6 distinctive jerseys for leading riders in such multi-stage races.  There does not seem to be a limit for team   “special bibs".  Thus, new "team value"  indicators could be implemented, thereby increasing the offer to new sponsorship. For example, by extension of the leading rider jersey notion, one could imagine that one defines a   “orange  bib for teams", similar to the distinctive  yellow bib, in Tour de France,  for the best time team ranking after $s$ stages, - the latter ranking not distinguishing if  riders accounting for some $A_s$ have abandoned the race.
  
Finally, one can also imagine that the $A_s$  indicator can be of interest for betting schemes (Yüce, 2021)
 and/or e-gamers (Beliën et al., 2011). 
 
 
    \bigskip 
 ==================== ==================== 
   
     \bigskip 
{\bf  Acknowledgements :}
  \bigskip 
 
   {\it to conserve anonymity, during the peer review process, no acknowledgement is hereby presented; it will await publication time ; reviewers, editor, and private communication expert will be mentioned }
  
{\bf   Data availability} : {\it data is freely available,  see text.}

    
  {\bf   Funding} : {\it this information will await publication time ; }
    
      
      {\bf  Disclosure Statement  on competing interest} : {\it Neither  relevant financial nor non-financial  competing interest has to be mentioned.}
      
      The author claims that "there is no conflict of interest".

 ==================== ==================== 
 
  \clearpage
   {\bf  Appendix    A: Disqualification Effects}
        
      \vskip0.5cm
      If  the question "Can one substitute riders during a multi-stage race?" may have some relevance, the complementary question "Can one remove riders during or after a race?" maybe studied within the present framework.
      
Indeed, it has been recalled that the 1999 and 2022 TdF final time  results and riders classifications have been later modified because riders, L. Armstrong and N. Quitana, respectively, were disqualified. 
In fact, these disqualifications  imply some modification of the team rankings as well.  The effect of such a complementary decision is hereby studied.  Since the data in the main text excludes both riders, "re-substituting"  them into the final riders classification leads to a lowering of their team final time.


Practically, the 1999 TdF $T_L$ result has to be modified as follows in order to re-substitute L. Armstrong:
the recorded time of 
L. Armstrong (USP), after the $L=20$ stage, was
 $ t^{(USP)}_{1,20}$ =  91:32:16; it had been
replaced by that of  the 4-th fastest USP rider,
F. Andreu. The latter rider original ranking place was 65; his final arrival time was
$ t^{(USP)}_{65,20}$ =  93:31:17.
Thus Andreu had arrived 1:59:01 behind the winner, L. Armstrong.
Therefore, substituting Andreu by Armstrong, the USP team final time has to be decreasing from 
 $T_L^{(USP)}$ =  276:02:34 to $T_L^{(USP)}$ = 274:03:33.
 Similarly, $A_L^{(USP)}$ =  278:08:52 becomes
 $A_L^{(USP)}$ =  276:09:51.  

 In so doing,  {\color{black}(it seems fair to emphasize that)} the USP team goes up  from the 8-th to the 1-st rank and from the 12-th to the 3-rd rank in the $T_L$ and $A_L$ classification respectively.
  
   

  In the case of the 2022 TdF, the "without Quintana" results are  modified  to those "with Quintana" as follows:
N. Quintana  (ARK) final time was  $ t^{(\#)}_{6,21}$ =  79:49:59, excluding  6 sec bonus; he  "had arrived" 6-th, 0:15:57 (not taking into account time bonus) after the 2022 TdF winner, J. Vingegaard. The latter finished the race in 79:33:52, excluding 32 sec bonus. The Quintana data has to  replace that of  his teammate C. Swift
(original ranking place 70),  $ t^{(ARK)}_{70,21}$ = 83:08:25.
   Swift "had arrived" \underline{ 3:18:26 behind his ARK leader} (excluding Quintana time bonus); 
Swift had arrived 3:34:33 behind the TdF winner (excluding time bonus). 
Thus, the team final times (after the $L=21$ stage) decrease  from
 $T_L^{(ARK)}$ = 242:59:54  and  $A_L^{(ARK)}$ =  247:44:49
 toward
 $T_L^{(ARK)}$ =  239:41:28 
 and $A_L^{(ARK)}$ =  244:26:23,
 under "rider substitution".  

 In so doing,  {\color{black}(it seems fair to emphasize that)} the ARK team  ranking goes  up from the 9-th to the 3-rd rank in the $T_L$ and  from  the 15-th to the 11-th rank in the $A_L$ classification.


In order to observe wether the substitution (due to the fastest rider removal)  has some effect, one can calculate the Kendall $\tau$ rank-rank correlations  between  $T_L^{(\#)}$ and $A_L^{(\#)}$, -  i.e., taking into account disqualified riders or not; see  Table \ref{Table5KtauTdFGdIwno}.  The rank-rank correlations are seen to  increase when taking into account the disqualified riders.  {\color{black}(It seems fair to emphasize that)} this is because such riders are in the top of the final classification. One observes  more homogeneous distributions and more regularity in the clustering.

    \begin{table}\begin{center}  
\begin{tabular}{|c||c|c||c|c|} 
 \hline 
&\multicolumn{4}{|c|}{TdF} 
  \\	\hline
 &	$1999$ &	$1999$ 	&	$2022$	& $2022$ 
   \\ 
  \hline	\hline
$N_L$   &	$140$ &	$141$ 	&	$134$ &$135$	
   \\ 
  \hline	\hline	 
Kendall-$\tau$	&	 0.75789	&0.77895 &	0.80952&  0.82684	 
 \\	
2-sided p-value	&	 0.00	&	0.00	&	0.00	&	 0.00	
 \\	
Score	&	 144 	&	148 	&	187&	191	  
 \\	
Denominator	&	 190	&	190	&231	&231		 	
 \\ \hline	 	
\end{tabular}
   \caption{ Illustration of Kendall $\tau$ rank-rank correlation  results between  $T_L^{(\#)}$ and $A_L^{(\#)}$ corresponding to the multi-stage races hereby examined, - taking into account disqualified riders or not in 1999 and 2022 TdF; $N_L$: number of  taken into account  finishing riders.
}\label{Table5KtauTdFGdIwno}
 \end{center}
 \end{table}

 \end{document}